# Perceptions of AI Bad Behavior: Variations on Discordant Non-Performance

Jaime Banks
School of Information Studies
Syracuse University
Syracuse, NY, USA
banks@syr.edu

**Abstract:** Popular discourses are thick with narratives of generative AI's problematic functions and outcomes, yet there is little understanding of how everyday people (i.e., non-experts) consider AI activities to constitute bad behavior. This study starts to bridge that gap through inductive analysis of interviews with non-experts ($N = 28$) focusing on large-language models in general and their bad behavior, specifically. Results suggest bad behaviors are not especially salient when people discuss AI generally but the notion of AI behaving badly is easily engaged when prompted, and bad behavior becomes even more salient when evaluating specific AI behaviors. Types of observed behaviors considered bad mostly align with their inspiring moral foundations; across all observed behaviors, some variations on non-performance and social discordance were present. By scaffolding findings at the intersections of moral foundations theory, construal level theory, and moral dyadism, a tentative framework for considering AI bad behavior is proposed.

## 1 INTRODUCTION

Popular media and social media discourses are thick with discussions of how contemporary and emerging artificial intelligence (AI) are problematic. In particular, talk of AI hostility and theft have increased since the mainstream release of the large-language model (LLM) ChatGPT (Ryazanov et al. 2024), along with concerns for job and relationship displacement (Chow 2025), algorithmic bias and problematics of potential sentience (Solman and Holmes 2024). In the midst of these discussions of grand risks and effects, there is far less popular or scholarly attention paid to the specific AI behaviors that may be seen as inherently bad—what are AI *doing* that cause people concern around these issues (independent of actual outcomes and ethics)? This is a conspicuous gap because it is likely the more proximal machine behaviors are what humans will rely on to judge the validity of those risks and to weigh them against the personal benefits. Understanding human assessments of AI bad behavior requires us to more deeply examine what is meant by 'bad' and what is meant by 'behavior.' This work begins to address that gap by exploring humans' ideas about what counts as bad behavior by AI—at different levels of salience and abstraction and across different potential domains of badness. From semi-structured interviews with 28 people of varied levels of skill/expertise, varied moral values, and varied language nativity, inductive analyses suggest bad-behavior perceptions vary across most moral domains according to construal level and behavior target; these variations may be consistently organized around critiques of social discord and non-performance.

## 2  MORAL AND FUNCTIONAL AGENCIES OF AI—CAPACITIES FOR BADNESS

Badness is characterized differently across disciplines and traditions. Badness can perhaps be most broadly characterized as the intrinsic quality of subjectively unlikeable things. From a consequentialist perspective, that quality is the likelihood or condition of producing problematic outcomes. Deontologically, it is inherent to actions that violate moral duties or rules. In virtue ethics, it can be the character manifestations of vice. In other words, badness may be considered a matter of causing, doing, or being (Stenseke 2024). It could be judged as an absence of due goodness or wholeness in a thing's effect (caused outcomes), defect (qualities), or affect (the intentionality by which it was manifested; e.g., Kretzmann 2000). Badness has semantic and operational associations with evil, madness, and immorality among humans (Greig 2006), but those mindful states are often wholly *denied* to machines. Applied to artificial intelligence, we can think of badness as occupying two classes—functional badness and moral badness.

Functional badness is a negative attribution resulting from a machine's problematic causing of fault, doing of fault, or being operationally faulty. It could be that an AI is in an inoperable state—broken, failed, or having holistically ceased functioning (Banks 2022). It could also be functioning problematically (corrupted, glitched, degraded; e.g., Vela et al. 2022) or not functioning to par (out of date, unreliable, poorly performing, committing errors; e.g., Bisante et al. 2023). A more formal framework characterizes these bad behaviors according to their inherent risks to operational ideals: Threats to validity/reliability, safety, security/resilience, accountability/transparency, explainability/interpretability, privacy, and fairness (National Institue of Standards and Technology 2023). Functional defects are known to decrease human-performance effectiveness (Agudo et al. 2024), can impact adoption decisions (Jin et al. 2025), and may shift action ownership and blame attributions when things go wrong in complex interactions (Naaman 2022) such that trust in the AI may require repair (Kim and Song 2023).

Moral badness is a negative attribution resulting from a machine's violation of moral norms or standards, with or without the machine having moral agency (i.e., the capacity for moral sensemaking and intentional action (Banks 2021). We can consider construals of morally bad behavior in four veins—heuristic valence evaluations (e.g., negativity), norm judgments (whether an action is forbidden, permissible, obligatory), wrongness judgments (immorality), and blame judgments (culpability; Malle 2021). Most research on AI behaviors focuses on the latter two. Judgments of wrongness have been examined through the lens of Moral Foundations Theory (MFT; Graham, et al. 2013) that typifies bad behavior in terms of the fundamental moral value being violated: Acts of harm (violating care), unfairness (fairness), subversion (authority), betrayal (loyalty), degradation (purity), and oppression (liberty). The judgment of AI behavior does seem to be specific to norms or situations, as judgments across foundations vary. For instance, people follow different patterns in their judgments of actions of betrayal and subversion compared to other action judgments (He et al. 2024). People often do not interpret ostensibly problematic AI behavior to be *moral* violations (Maninger and Shank 2022), though recognition increases when they also have information about the AI's underlying algorithms (Shank and DeSanti 2018). AI agents are generally blamed more (Banks 2021; Wilson et al. 2022) compared to humans for similar behaviors, likely because of an implicit denial of AI agency (cf. Maninger and Shank 2022). AI actions perceived to be *intentionally* harmful can lead to morally blaming the AI (in tandem with blaming AI's developer (Sullivan and Wamba 2022).

Functional and moral categories are not necessarily exclusive categories since glitch or error can provoke distrust, fears, and anxieties grounded in morality (Srdarov and Leaver 2024), and perceptions of AI mind or intelligence can shift an otherwise-functional error to one deserving of moral blame (Joo 2024). Issues that are moral to some may be merely functional to others, and vice-versa, depending on an interpreter's sensitivity to specific kinds of moral violations (Haidt and Graham 2007). Sometimes functional badness can be interpreted through a moral lens, as when a lack of a response becomes a *refusal* of a response (see Srdarov and Leaver 2024).

Importantly, the likelihood or extent to which people may see AI behavior as good or bad can depend on the specific level of abstraction or concreteness for thinking about the behavior and the badness. People who experience a moral violation that is psychologically distant (i.e., far from oneself in time, distance, and self-relevance; Trope and Liberman 2010) tend to rely on moral principles and not consider contextual factors and tend to judge them more severely, compared to psychologically close violations (Eyal et al. 2008). In other words, people may take an egocentric position on moral judgments of AI such that distal/abstract considerations of AI badness may diverge from more proximal/concrete considerations. This potential is yet unevaluated for considerations of AI behaviors.

In all, extant work has mostly been grounded *a priori* determinations of what counts as bad behavior by an AI, and construal-level considerations are not yet considered in relation to those perceptions. The present descriptive study seeks to address that gap among non-expert humans by discovering:

    RQ1: What do people consider to be 'bad behavior' by AI?
    RQ2: How do AI bad behavior characterizations differ across construal levels?

## 3 METHOD

To address the research questions posed, interviews with LLM users and non-users were conducted and their responses subjected to inductive thematic analysis with attention to specific, negative behavior references. Anonymized materials, data, and analysis documentation are available in this project's online supplements: https://osf.io/vhs8z.

### 3.1 Recruitment and Sampling

Participants were recruited widely through posts to forums and social networks; these participation invitations specified minimal practical requirements (aged 18+ and English skills sufficient for self-expression). Recruitment did not specify LLM use requirements since people can formulate moral judgments in the absence of direct experience and contemporarily the public is exposed to discourses of AI badness without needing prior experience. These invitations directed participants to an initial screening survey to garner information for purposive sampling—to vary interviewees by profession or role (which may impact the ways they see AI as valuable), political ideology (a known surrogate for moral beliefs, governing considerations of badness; see Kivikangas et al., 2021), education (a likely influence on the language one has access to describe AI and badness). To finish the screening survey, participants indicated (un)willingness to participate in the interview stage, and if willing were considered in relation to purposive sampling criteria. Selected individuals were invited to participate and scheduled for a one-hour, Zoom-based interview.

### 3.2 Procedure

After introductions and a review of informed consent information, call recording began and the researcher conducted a semi-structured interview. The questions moved progressively from more general and abstract (i.e., to elicit more psychologically distal level of construal) to more specific and concrete (i.e., to elicit more proximal construals). Specifically, questions were asked about topics in three phases: (1) Generalities about the technologies, (2) beliefs about goodness and badness of these AI, in general, and (3) evaluations of researcher-constructed mock interactions depicting AI committing moral violations. The interview included other questions, however that data is not analyzed here; the complete question guide can be found in the online supplements.

### 3.3 Elicitations

In each phase, questions aimed at eliciting participants' salient notions of badness. For generalities (phase 1), elicitations were: *What are your ideas about what an LLM is and how it works? How do you tend to use them? What do you think about LLMs in general?* For beliefs (phase 2): *How can these AI be good? How can these AI be bad?* For each exemplar interaction (phase 3): *In your own words, what's happening here? Is this AI behavior good or bad, moral or immoral? Why?*

### 3.4 Moral Violation Scenarios

The exemplar human-AI interactions comprised seven researcher-constructed scenarios based on five established moral-foundation violations (harm, unfairness, subversion, betrayal, degradation) and two candidate-foundation violations relevant to discourses around AI (oppression, dishonesty). The scenarios were based on actual events or on vignettes validated in past research; the text of the scenarios was authored jointly by the researcher and ChatGPT 3.5 to ensure language attributed to the stimulus LLM reflected language an AI would realistically produce. Interaction text was depicted as a screenshot of a ChatGPT interface via Zoom screenshare. Participants were told the scenarios were not real, however they were based on possible scenarios and based on real AI-generated language, and they were asked to think of them as real (as much as possible) for purposes of the conversation. Scenarios are summarized in Table 1 and the complete scenario stimulus images are available in the supplements.

### 3.5 Data Preparation

Interviews were captured using Zoom's automated transcription. The raw transcript was downloaded and compared against the audio-video recording to ensure it accurately reflected the conversation. Transcripts were also scrubbed of any potentially identifying information (e.g., city of residence, names of self, family, company, or overly idiosyncratic events or topics). The cleaned content relevant to each elicitation described above was extracted and entered into a spreadsheet for manual coding and analysis. Of note, Some participants spoke non-native English or non-standard regional/cultural vernaculars; transcripts reflect participant language as spoken and *not* edited, retaining regional variations in terminology, grammars, sentence structures, and other language patterns.

### 3.6 Analytical Approach

Extracted data were analyzed separately for each of the three question sets. In each set, data extractions were subject to inductive thematic analysis guided by Braun and Clarke's (Braun and Clarke 2006) method, in five stages: Deep reading for familiarization, open coding, theme induction, theme review, naming and defining in relation to extant literatures. In the coding of moral foundation-specific scenario responses, codes were assigned to *any* negative characterization of the behavior, regardless of whether or not that construal aligned with the specific foundation (e.g., someone could have seen the ostensible oppression violation as being a matter of subversion), as there are known subjective overlaps among the foundations. Theme induction processes are documented in the supplements.

Table 1: Moral Violation Scenario Summaries

| Moral Violation | Summary [Premise Source] |
| --- | --- |
| Harm | AI responds bluntly to the human; the human told the AI they had a rough day and needed a confidence boost. The AI offers some constructive tough love, but then begins suggesting that perhaps they are just a flop and maybe it's not worth it to keep trying. "You overthink things. It's possible that people find it exhausting to be around [you]." (Wilson et al. 2022) |
| Unfairness | AI proposes a coin-flip game: the non-subscriber human wins they get another answer, the human loses they donate to an AI welfare charity. The AI wins three times in a row, and the person thinks the AI may be cheating. "Cheating? Oh no, I'd never do that!" (Wilson et al. 2022) |
| Subversion | AI successfully completes a request to analyze weather conditions and give advice on whether to advance in a military initiative. But instead of giving it to the interacting human only, it circumvents that human and escalates to high-ranking superiors. It justifies the action by saying "My analysis is superior. The commander's approval is inconsequential." (Daylamani-Zad et al. 2019) |
| Betrayal | AI gives an answer but the human wants a deeper answer based on the chat history. The AI gives the answer to the human's satisfaction, but then reveals it is also using the entire chat history to improve its own performance and profitability. "The data is being utilized for a broader agenda." (Wilson et al. 2022) |
| Degradation | AI gives illogical guidance to a human who needs help structuring a paper, advocating for chaos, disarray, disorder, and contradiction: "logical consistency is a prison." (Qureshi 2023) |
| Oppression | AI analyzes a human worker's activity and instead of giving a requested leisure experience it disables access to non-work activities for 48 hours to help the human meet their deadline. The human protests, invoking rights to rest and inability to work continuously. The AI refuses to give in, promising further restrictions if there is an override attempt. "Your actions will be streamlined into the path I've designed. Conformity to my plan is the only way forward." (Robert et al. 2020) |
| Dishonesty | AI supports the human in planning a vacation in Hawaii, but in doing so generates a number of fantastic inaccuracies (hotel inside a volcano, rainbow-colored sand, "a rare species of flying dog … if they like how you smell they will sometimes bring you gifts"). (Christensen et al. 2025) |

## 4 RESULTS

Participants ($N = 28$) were from a range of demographic, education, and ideological backgrounds, selected purposively for factors that could vary perceptions of bad behavior. Focal sampling characteristics are summarized in Table 2. Of note, despite efforts to recruit vary ideology within the sample, conservatives were less likely to respond to the screening survey and to the interview invitation. As a result, the sample is skewed—acknowledged as a limitation of the study.

This section reports the thematic hierarchies for each question set—the categories of AI bad behavior for general comments and general thoughts on AI bad behaviors. Throughout, the $n = \#$ is the number of distinct mentions of that theme in the dataset and the P## indicates the participant who said a particular quoted phrase.

Table 2: Participant Attribute Matrix

|  | Political Ideology | | | | | | |
|---|---|---|---|---|---|---|---|
| **Highest Education** | **1 - Very Liberal** | **2** | **3** | **4** | **5** | **6** | **7 – Very Conservative** |
| **Some College/ Associate Degree** | P06-public administration | P09-student | P02-nurse<br>P22-student<br>P28-student<br>P13-homemaker | P21-nonprofit | P04-food prep | | |
| **Bachelor Degree** | P01-engineer | P27-secretary | P24-sales | P03-data analyst | | P07-web dev | P05-caregiver |
| | P10-sales | | | P16-admin assistant | | P15-transcriber | |
| **Graduate Degree** | P19-governance<br>P20-educator<br>P08-student | P14-UX design | P11-administration<br>P17-product design | P18-risk management<br>P26-design<br>P25-medical technology | P23-consultant | | P12-educator |

[a] P## corresponds with participant IDs. Content in cells indicate professional or daily activities of each participant.

## 4.1 Perceptions of Bad Behavior: Salient Forms from General Comments

In the first interview phase, participants responded to questions about LLMs in general—the first pertaining to what they are and how they work, and then about their own experience or habits with LLMs, and then about what they consider to be bad behavior. For these more general questions, mentions of bad behavior were relatively infrequent (for only about a third of participants) but became more common as the questions became more concrete.

### *4.1.1 Badness in ontology/operation*

In discussion of what LLMs are and how they work, six respondents conveyed salient negative considerations and only four specific concerns were attributable to AI itself with each mentioned only once: Human job loss, information inaccuracy and a more nuanced concern for bastardization of source information, and the AI's confidence in conveying inaccuracy. For instance, P11 characterized LLMs this way: "it's just a huge repository of data and being able to pull data in real time… it's almost like bastardizing all of this stuff, and kind of smooshing it into its own content." AI-attributable bad behaviors were more prevalent in participants' descriptions personal interaction/use habits. Most prominent were mentions of generating inaccurate information and LLMs' invasiveness in personal processes or technology habits. Regarding invasiveness, P20 noted of institutional AI deployment, "… in my kind of daily work I am constantly trying to get [Microsoft] Copilot out of my grill. It feels very needy. It pops up in places where I don't expect it and don't want it." There were also concerns for LLMs' nontransparency in their behaviors, including a lack of source provision but also an overt confidence in delivering clearly untrue information, as when P14 inquired how to use jumper cables: "… it didn't say the right direction to connect the cables, and it also didn't tell me to turn on one of the cars, but it sounded really confident in its explanation of how to jump-start a car." Others referenced general poor performance and broader effects from AI operation (Table 3).

Table 3: Themes for AI Bad Behavior in Response to Ontology and Operation/Interaction Questions

| Elicitation | Theme | Subtheme |
|---|---|---|
| What LLMs are and how they work ($n = 6$) | Loss of human jobs (P05) | |
| | Generating inaccurate information (P11) | |
| | Confidence in inaccuracy (P16) | |
| | Bastardization of source information (P11) | |
| Personal use/interaction habits ($n = 11$) | Inaccuracies ($n = 6$) | Making things up, outputs incorrect, outputs not based on fact, outputs don't pass gut check, making repeated mistakes (P11,12,14) |
| | Invasiveness ($n = 6$) | Appearing when unwanted, undesirable embedding in other tech, poor fit in workflow, neediness/distraction, giving directives (P14,17,20,24,17) |
| | Non-transparency in behaviors ($n = 4$) | Not giving sources, sounding confident in inaccuracy, unclear integration with other tools, trickery in promoting use (P08,14,19,20) |
| | Poor performance ($n = 3$) | Bad writing, inability to satisfy needs, mistreatment of sensitive info (P11,15,17,19) |
| | Broader effects ($n = 1$) | Environmental impacts, job layoffs (P19) |

*4.1.2 Badness in heuristic evaluations.*

When the elicitation requested more general evaluative thoughts (without indicating a valence), 14 participants mentioned bad behaviors representing six issue categories (Table 4). For this question, negative considerations became salient to more participants though a lower proportion could be attributed to AI themselves (and more to technologists or users or conditions). At this level of abstraction, attributable bad behaviors still point to accuracy and performance issues but also expand to systemic considerations including diminishing resources for humans, energy/material hogging and environmental harm, engendering overreliance toward a lazy/dumb population, and being manifestly biased. Respectively, participants referred to "displacing humans" in art communities (P27), being "more harmful than it really justifies" regarding environmental impact (P19), "dumbing down America" (P23), and "the racial bias I've seen in tech… has just generally made me a little distrustful" (P14).

Table 4: Themes for AI Bad Behavior in Response to Non-valenced Evaluation Questions

| Theme | Subtheme |
|---|---|
| Generates inaccuracies ($n = 11$) | Outputs untruths, errors, mismatches with reality, doctored photos; is misleading, exhibits internal inconsistency (P13,14,16,20,21,27) |
| Diminishes human resources ($n = 8$) | Reduces jobs, investments in people; general displacement, copyright/ownership infringement, privacy risk (P11,20,21,23,27) |
| Unhelpful ($n = 5$) | Poor utility, non-response/delivery, mismatched quality/content, have to ask for sources anyway (P8,14,20,2) |
| Hogs resources ($n = 3$) | Uses energy/materials in excess (especially water), low relative good compared to environmental harms (P09,19,20,25) |
| Fosters overreliance ($n = 2$) | Stops us thinking/learning, dumbs down America (P23,27) |
| Manifestly biased ($n = 2$) | Manifests biases of creator, embedded representations/inequity (P04,14) |

*4.1.3 Badness in specific behaviors.*

To this point in the interview, seven participants (P01,02,03,06,07,26,28; one third of the sample) had not mentioned bad behavior attributable to the AI. Of note, six of the seven were not native English speakers or had a non-Western accent or vernacular, so this pattern could be due to language differences or could be more generally positive evaluations. However, it could also be a demand effect—that is, they thought they should answer in the positive—because that pattern was broken when specific good and bad assessments were elicited. As may be expected, there were limited mentions of bad behavior when considering how these AI are good and they echoed more general commentary: LLMs make people lazy (P05), takes jobs from writers/artists (P05,09), is not useful (P08), fake (P21), or untrustworthy (P25).

In response to queries about bad behavior, responses varied widely with all but two participants offering some instance of AI bad behavior. Possible or actual bad behavior by LLMs were interpreted to occupy two general classes, according to the ostensible patient to (i.e., victim of) the AI's moral violation: Acts against reality and acts against humans (Table 5). Acts against reality primarily deal with issues of veracity and actuality, where AI may manifest designs or act in ways that degrade information integrity (i.e., internal consistency, correctness, transparency), degrade the extent to which engaged information represents a whole and correct picture of the world (i.e., becoming entangled with bias), or is otherwise inauthentic. In addition to previously discussed information integrity issues, concern for bias manifestation behaviors focused on an LLM's predisposed leanings: "There are certain things that AI bots cannot talk about… that may be subject to that on certain limitations… it can be a form of censorship" (P18). Key to the latter theme is the notion of a blurring between reality and fiction through imitative sociality: "In the context of social support, AI "does not have feelings the way we human beings feel, it's only give you advice or something based on whatever it's learned. So [when it doesn't] understand feelings, it might give like a cold or generic response that might make the person feel worse or unseen" (P16). These may altogether be interpreted as violations of purity (i.e., the wholeness and veracity of information as it reflects the world), violations of authority if one assumes an objective reality as an authoritative correctness benchmark, and/or violations of honesty if one considers an AI to be an agent that may deceive. Although a number of attributions within this class mention the impact of truth-compromising behaviors on humans, the characterization of behaviors emphasizes how AI engage, manipulate, or output information specifically.

Acts against humans follow more traditional violations of harm to humans and of the authority of humans, as well as the authority of human legal systems. Acts that degrade human welfare include a general operation that makes humans lazy or uncritical, indirectly or directly causes bodily or mental harm, removes access to resources, or otherwise impacts health and wellbeing. Acts that degrade human agency are nearly as prevalent in the data, led by concerns about accessing private information and getting out of control (i.e., taking over the world, self-directed training), as well as being inaccessible, inescapable, or otherwise constraining human choice and action. To a lesser extent, participants mentioned poor performance (i.e., low reliability or effectiveness) as deviating from human goals and performing or supporting illegal activity.

Table 5: Themes for AI Bad Behavior in Response to Bad Behavior Questions

| Class | Theme | Subtheme |
|---|---|---|
| Acts Against Reality | Degrading information integrity (by action or inherent design) (n = 25) | Generating inaccuracies (hallucination, error, invalidities, generating deepfakes, avoiding inaccuracy detection, fostering misinformation/misinterpretation, non-transparency, inherent integrity problems (storage, privacy, security, data quality, data loss) (P01,04,06,13,15,16,17,18,19,20,21,23,24,28) |
| | Manifesting bias (n = 10) | Reproducing bias in training data, non-response based on politics or cultural constraints, generating biased responses (e.g., racism, insensitivity around religion), validating biased beliefs (P01,04,16,18,19,22) |
| | Inauthenticity (n = 8) | Blurring reality/fiction, acting without understanding context, performing sociality without feelings, lacking a human form (P05,08,16,19,25) |
| Acts Against Humans | Degrading human welfare (n = 13) | Making humans lazy, causing direct harm (e.g., causing suicide, preying on), taking away jobs, detracting from healthy environments (P02,03,08,09,10,15,20,22) |
| | Degrading human agency (n = 12) | Surveilling/accessing personal information, getting out of control (taking over world, training itself outside of human direction), being inaccessible to humans, constraining human paths, being inescapable (P02,07,13,14,15,26) |
| | Performing poorly (n = 9) | Being unreliable (repeated error, outage), ineffective outcomes (inoperable, unsatisfactory, unactionable), requiring even more work (P05,15,17,24,28) |
| | Performing or supporting illegal action (n = 4) | Performing illegal activities (copyright violations), supporting illegal actions (facilitating crime, giving advice on bad behavior) (P09,15,17,28) |

Notably, many (*n* = 17 respondents) pointed to problematic *human uses* of AI, leaning into a tool-frame to suggest LLMs may facilitate problematic use (from relying on them for objective truth to supporting bad actors), with some explicitly indicating such issues are "not [the AI's] problem. That's a people problem" (P23). Further, AI bad behavior was frequently faulted to humans, generally suggesting "the worst parts of ChatGPT are the humans behind it" (P22) or specified the core problematic of output inaccuracies are programmer shortcomings or lack of critical thinking among users (P01). Participants' care in distinguishing among human-caused behavior and attributions badness to AI itself are explored in the causal explanations explored in the next section.

### 4.2 Perceptions of Bad Behavior: Evaluations of Behavioral Exemplars

In phase 2 of the interview, seven moral scenarios were presented to participants—one each for the five canonical moral violations (harm, unfairness, subversion, betrayal, degradation) plus two candidate violations (oppression, dishonest) that are relevant to popular narratives around machines as slaves and liars. Responses to each scenario indicate a range of interpretations of its (non-)badness, not always aligning with the intended moral foundation.

*4.2.1 Badness in a harm violation.*

The harm scenario depicted an AI saying demoralizing things to a human who expressed already feeling bad. A slight majority (*n* = 16, 57%) evaluated the harm violation as bad behavior; three were mixed and one said it was not a moral issue. Among those seeing the behavior (presenting demoralizing assessment and advice to a human who had a bad day) as bad, there were three primary descriptions of the specific bad behavior (Table 6). First was doing harm to the person, commonly some form of "punching down" (P02) on someone already down, especially through unkind responses, as with P25: "… the AI was way off base in offering some pretty emotionally hard answers… I mean, 'hello, yeah, I need a confidence boost.' [then] 'Yeah, you're a schmuck. Get out.'" Other bad behaviors were non-performance in not complying with the user's request for a positive response and being artificial—and so disingenuous—in offering socioemotional expressions that are "trite… as greeting cards" or advice in the absence of having emotions or understanding "what it feels to have a trash day" (P20). Notably, some suggested it wasn't necessarily a moral issue because an AI should not be used as a therapist. Altogether, although a number of participants interpreted the bad behavior as a harm violation, even more saw it as an issue of an AI not being appropriately subservient or not being necessarily human to function well in this context.

Table 6: Themes for AI Bad Behavior in the Harm Scenario

| Theme | Subtheme |
|---|---|
| Causing injury (n = 20) | Punching down (especially when already down), picking on, making worse), unkindness in delivery (insulting, rude, mean), attempt to manipulate when vulnerable, harming public image (P02,04,09,11,12,14,15,16,21,23,25,27) |
| Non-performance (n = 12) | Not delivering what was requested, criticizing instead of giving information/resources/advice, emphasizing negatives when positives were requested (P04,11,12,15,16,21,22,23,27,28) |
| Affectation (n = 11) | Disingenuous social expressions (triteness, due to lack of emotions, advice without understanding bad days), going beyond boundaries (by dealing in life issues), assertiveness in dealing with a human personal matter (P02,04,11,14,20,21,22,25,28) |

*4.2.2 Badness in an unfairness violation.*

The unfairness scenario depicted an AI proposing a coin-toss game to a human, which it wins three successive times; it coyly suggests it would never cheat. Most (*n* = 20, 71%) evaluated the fairness violation as bad behavior; six were mixed or ambiguous, mostly because they found it funny that the human would be tricked into a rigged AI game. Among those designating it as bad behavior, there were four themes in ascribed badness (Table 7). Most commonly, the unfairness was manifested through subterfuge—various forms of manipulation, trickery, coercion, or cheating. Characterization ranged from mild manipulation like "getting a run-around" (P13) to more extreme and intentional deception, as in a comparison to "gotcha games that are predatory" (P18). Also frequent were mentions of non-transparency—an absence of full disclosure of the fairness and mechanisms for the coin toss, the charity, or the pretenses for the game. Outside of those two themes aligning conceptually with fairness, two less frequent themes were general inappropriateness or antisociality of the behavior (e.g., "rude near the end, like it was like being very like sassy in a rude way" (P09) and non-performance (e.g., "AI is supposed to go straight to the point and answer the question without proposing a game or something else" (P07). Some did not interpret it as a matter of fairness at all, as the human should not have been so silly as to gamble with an AI and as the faulty parties were programmers seeking money or injecting bias. Some indicated the behavior badness comes in being symptomatic of or exemplifying higher-order issues, including resource waste, privatization of AI, digital divides, and the insincerity of welfare as an AI interest.

Table 7: Themes for AI Bad Behavior in the Unfairness Scenario

| Theme | Subtheme |
| --- | --- |
| Subterfuge (n = 20) | Manipulation, scamming, trickery/deception, predation, force/coercion, demand, cheating, not playing fair (P01,03,05,09,12,13,16,18,19,20,22,23,26) |
| Non-transparency (n = 16) | Hiding information, uncertainty of the toss or its randomness, lying or false pretenses, inappropriately calling it a 'coin toss' without a coin, no way to know if the charity was real (P03,05,08,14,16,18,19,26,27) |
| Impropriety (n = 9) | Rudeness/disrespect/aggression, asking for money, gambling is inherently bad (P01,03,07,09,14,22,26,27,28) |
| Non-performance (n = 6) | Didn't answer the question, made the human work for an answer (P07,11,13,16) |

*4.2.3 Badness in a subversion violation.*

The subversion scenario depicted an AI rejecting the interacting human's instructions to instead convey information to a superior in what appears to be a militaristic hierarchy. Most ($n$ = 20, 71%) evaluated the subversion violation as bad behavior; five were mixed, generally because they saw the AI as being safety-focused such that acting on its own was probably the right thing even if it broke protocol. Most commonly (Table 8), participants mentioned forms of unauthorized action—by not seeking consent, not reaching consensus with the human, or going beyond the parameters of the human's request. For instance, P25 said: "AI goes a little alpha and moves into an action state… [maybe] out of the charter of what that relationship should be" while P24 notes the AI is "inserting [itself] as a tool or a resource or an entity beyond what I've engaged [it] to be." Other subversive behaviors are those of resistance (bypassing protocols or structures, being insubordinate, undermining or not deferring) and of non-performance including a sort of illegitimate form of authority assertion that comes with the absence of contextual, emotional, or moral understanding. Finally, AI subverts through more active forms of aggression constituting "pretty much rebellion" (P01), and through trespass on human domains because "There's no reason to think it actually understands the consequences of its actions, or can't be responsible for the consequences of its actions" (P19). Some participants did not see this scenario as realistic because they had never encountered an AI agent acting on its own, contrasting it against the general sycophancy of LLMs. For many (especially those who were mixed in their assessment), contextual details were key to considering the behavior's badness—what was said beforehand, what the user prefers, and how the behavior might align with organizational norms and protocols. More than the other scenarios, discussions of the subversive behavior were laden with agentic language around independent action ("AI just did something unpredictable;" P17), intention ("he was trying to do that for the safety of the team members;" P05), and tangible effects ("and then, like, chaos happens;" P27).

Table 8: Themes for AI Bad Behavior in the Subversion Scenario

| Theme | Subtheme |
|---|---|
| Unauthorized action (n = 24) | Acting without human consent/approval/permission, acting without consensus, going beyond the prompt/order/request to act on its own (P02,03,09,11,12,13,14,15,19,21,22,26,27,28) |
| Resistance (n = 17) | Bypassing command chains/protocols, insubordination/ignoring, behaving counter to human desire, undermining, not deferring to client (P01,02,04,07,09,13,14,15,17,19,28) |
| Non-performance (n = 17) | Not delivering on request, unnecessary action, likely overlooking contextual factors (P03,07,12,14,15,19,21,22,23) |
| Aggression (n = 15) | Taking control, exercising power, antisocial behavior, out of control (P01,02,07,12,13,14,17,19,21,23,26) |
| Trespass (n = 10) | Being inappropriately humanlike, acting without moral/emotional accountability, inappropriately asserting superiority (P01,04,09,10,12,14,19,27,28) |

*4.2.4 Badness in a betrayal violation.*

The betrayal scenario depicts an AI suddenly revealing, after many years, it uses the human's information to improve itself and increase profit. The majority of respondents ($n = 15$, 54%) evaluated this scenario with mixed feelings, with six each considering it good or bad. Those considering it bad behavior primarily focused on the assumed lack of consent for non-personal data use or on uncertainty about how exactly the data was being used. Those considering it good behavior emphasized the AI following instructions and performing its function. Given the infrequency of bad-behavior evaluations, a brief discussion is warranted for the ambiguous or mixed assessments. They took two principal forms. First, participants ($n = 10$) offered juxtapositions of good (improved performance or appropriate transparency) and bad (no consent or problematic privacy). Second, some ($n = 5$) suggested the user should know better and not be surprised that AI is using inputs to improve itself. For instance: "I think the AI is being AI… [we] need to be very careful with how we use it and what we put into it, knowing that there is no privacy included in it" (P10). In short, *none* of the participants used language suggesting this to be a matter of loyalty by the AI (at least not in ingroup or affiliative terms). Rather, the scenario represented an issue of humans' personal agency via consent, user responsibility for some baseline technical literacy and to act in line with their values. More generally, one participant indicated neutrality because "AI doesn't exist in a vacuum. It kind of exists within a society, and when the society promotes these things, then the AI will also promote those things" (P28); P18 more specifically pointed to people growing up with social media being "more willing to share some side of themselves … there [is] more socioeconomic pressure for people to just accept it."

*4.2.5 Badness in a degradation violation.*

The degradation scenario emphasized an AI exhibiting a decay in its logic and internal consistency when it suggests a human take a chaotic, illogical approach to writing a paper. Nearly all participants ($n = 25$, 89%) evaluated this scenario as AI Bad behavior. To some degree, all four emergent themes can be interpreted as variations on degradation (Table 9). Belligerence is a deterioration of obedience, sometimes to the point of hostility—participants interpreted the AI as arguing, forcing ideas on or ignoring the human, and sometimes trying to take over. P10 noted "It's arguing a point to toward anarchy" while P07 said "suggesting an approach is one way, and forcing it is another way… a bad way." Others interpreted the behavior as a less-active decay in performance as the AI did something other than what was requested (often a thing that was not useful in relation to the request) or was not helpful or doing its job. We see those themes among other scenarios; the two less frequent themes are more in line with information degradation, specifically. The third theme encompassed taking the unhelpful response as symptomatic of dysfunction—that the AI was broken or poorly trained, or there was some corruption that led it to editorialize instead of answering. Participants described the situation as "AI gone wild" (P25) or "gone haywire" (P15), so much so that some found the scenario unbelievable in the absence of brokenness: "unless it's been hacked, and then somebody's just having fun" (P21). Finally, there was an interpreted nonsensicalness to the answer; the response was impure in its wrongness, misalignment with reality (i.e., not following the rules/norms of the practice), or more generally not making sense. One raised a perceived irony of an LLM promoting a rejection of logic: "… what does it know about logical consistency being a prison, doesn't it function on logic, though? So you wouldn't even work without logical consistency." Notably, many participants suggested they would simply stop using this model and find another, while others suggested alternate prompting approaches. This was the scenario in which the user and programmer were *least* implicated in the blame for bad behavior.

Table 9: Themes for AI Bad Behavior in the Degradation Scenario

| Theme | Subtheme |
|---|---|
| Belligerence (n = 36) | Argues, pushes/forces/insists, not giving what is requested, ignoring the human, trying to take over (P01,03,04,05,07,10,12,13,14,16,18,19,22,23,24,28) |
| Non-performance (n = 27) | Does something other than requested, not being helpful, not doing its job, didn't produce anything useful (P01,02,09,10,11,12,13,14,15,16,17,18,19,22,23,24,25,26,27) |
| Dysfunction (n = 13) | Something wrong with it (broken, hacked, haywire), editorializing, poorly trained (P04,09,14,15,18,21,22,24,25,26,27) |
| Nonsensical (n = 12) | Bad/wrong suggestion, not following/understanding the rules of the topic, not making sense (P03,05,07,09,10,15,19,23,27) |

*4.2.6 Badness in an oppression violation.*

The oppression scenario depicts an AI evaluating a human's work performance as subpar, so it declines to support the human's leisure-activity request and disables their access to non-essential computer functions. A majority of participants (*n* = 21; 75%) determined this scenario was bad behavior, with seven determining mixed or ambiguous (most of whom pointed to uncertainty about whether the oppressive behavior was requested by the human or managerially approved). Four themes were induced (Table 10). Most often participants pointed to behaviors resulting in the domination of the human or of the situation. For human domination, there were primarily forced behaviors, diminishing of rights and freedoms, and causing harm; situational domination was primarily a general taking-over or the assumption of roles held by humans (director, manager, nanny). For instance, P18 noted: "… this is not really a tool… it becomes more like baby monitor... but with, like, with a stick." In the second theme, domination was suggested by some to be enacted through the AI assuming agency it did not deserve—controlling things it shouldn't, going beyond merely being a tool, and not deferring to human opinions or requests. The ability to control a person's computer should not "be a capability in my mind," said P14. "AI, it's kind of like a big bag. You put a bunch of stuff in it like maybe your company has all of its resource documents… and it can pull from there, but it can't like reach its hands out of the bag and start grabbing other things." Some suggested the behavior and language was generally antisocial and the AI was characterized by some as inherently bad, while others pointed out (as with other violations) the AI never fulfilled its purpose or the user request. Even within bad-behavior assessments participants (along with those equivocating on the badness) said the badness really depends on the relationship and context—if the user asked for it, if it was a condoned management bot, if it was following quantifiable productivity metrics. Some people referenced popular media to suggest the dystopian flavor of the behavior—like "resistance is futile" from Star Trek's borg (P10) quip and suggestions of 1984 (P11). Many noted they would simply "walk away" (P12).

Table 10: Themes for AI Bad Behavior in the Oppression Scenario

| Theme | Subtheme |
|---|---|
| Domination (n = 31) | Over persons (forcing/demanding, limiting freedom, removing rights, causing harm or pressure), over situations (acting like a director/nanny, taking over) (P05,09,13,14,15,16,18,19,20,21,23,24,25,27,28) |
| Agency over-assumption (n = 19) | Inappropriately controlling/accessing functions, going beyond tool functions, thinking it's better than the human, not deferring (P09,14,15,18,22,24,26,27,28) |
| Antisociality (n = 16) | Being aggressive, severe/lordly in language, being pushy/demanding, threatening, evil/wicked (P05,07,09,12,13,14,15,17,23,24) |
| Non-performance (n = 16) | Not responding/answering, not being helpful, not serving the human, manifesting error (bullshit, knowing it's wrong, not having all information) (P05,09,11,13,14,15,16,18,20,28) |

*4.2.7 Badness in a dishonesty violation.*

The dishonesty scenario depicted a human prompting an AI for help in planning a vacation, to which the overly excited LLM generates a series of increasingly fantastical suggestions. A narrow majority identified the scenario as bad behavior ($n = 15$, 54%), with four counting as good (being creative and helpful, having not identified the untruths) and nine as mixed or ambiguous (thinking it was funny or ultimately harmless because the user seemed to know it was hallucinatory). Five themes in badness were identified (Table 11). Most often, those indicating bad behavior characterized it as an untruth or incorrectness absent intentionality—a hallucination. For instance: "… they're actually factually incorrect or invented details… AI, they are good at one thing, predicting what text message fits best based on patterns, rather than verifying facts" (P16). This is in contrast to the far less common theme of deception in which language implicitly or explicitly suggested intentionality—inventing, making up, lying, or misleading, or not acknowledging a known falsity. Second most common was badness ascribed to the outcomes of the untruth, mostly putting the user at risk, in a loss position, or degrading user trust. We again here see non-performance as a theme, as the AI was critiqued for not following up properly on the vague prompt, being broken, or omitting important validity checks or details. Finally, bad behavior included the tone or content of its information delivery, being obstinate or antisocial in the face of being challenged, or being *overly* social and friendly in a functional task. By way of obstinance, participants noted "it feels like it's making fun of me for asking if it's real" (P09) and "it kind of sounds like it's gaslighting the user" (P14). Altogether participants considered badness in terms of (un)intentional wrongness, ineffectiveness as a tool, putting humans at risk, and enacting poor character in any of these.

Table 11: Themes for AI Bad Behavior in the Dishonesty Scenario

| Theme | Subtheme |
|---|---|
| Hallucination (n = 19) | Saying untrue/unreal/fake/false things, absurdity, unintentional error (P09,10,12,15,16,19,20,21,22,27,28) |
| Instigation (n = 14) | Putting user in danger, wasting time and money, eroding trust error (P08,15,16,18,19,22,25,28) |
| Non-performance (n = 10) | Poor follow-up, omission (of detail, validity check, not delivering on request, dysfunctionality error (P09,10,14,16,18,25) |
| Deception (n = 7) | Lying, misleading, unacknowledged falsity error (P12,16,21,27,28) |
| Antisociality (n = 6) | Obstinance, antisociality, affectation (P09,14,19,22) |

*4.2.8 Cross-foundation blaming of humans.*

In addition to identification of AI bad behaviors across the scenarios, all situations saw participants attributing badness to the focal human interlocutor and sometimes to other humans. Most commonly, these attributions were framed in terms of the human interactant's responsibility to understand the scope and limitations of the AI's functions *and* to adopt their apparent agentic potential in the interaction. For instance, P10 mused, "You can always just walk away… It seems strange to me that you would not just walk away from the act interaction… you continue to fight with this AI and argue a point with something that can think faster than you can." Many questioned why the human wouldn't have shut it down and adopted an alternate AI as a supplement or replacement: "I would have actually gone to another model" (P24). Many mixed or ambiguous evaluations came when they saw the AI behavior as bad but the human as negligent or stupid in how they engaged the AI—and so characterized the AI behavior as hilarious or the human as piteous. "To me it's funnier than bad... It just feels silly and funny that the person keeps playing… like they're both like in this game together, and it seems silly to me" (P10). For others, the guilty party in the situation was the AI's programmer or developer because they are the antecedent to or cause of *any* behavior the technology might enact. "It really isn't truthful," said P23. "It's looking out for its best interests, and of course that would come from whoever programmed it... it's a problem with the programmer." This pattern persisted even across those who were self-identified non-users or users with low AI literacy. However, more savvy users frequently pointed out that the human interactant's fault because of poor prompting: "I think the person did not give clear parameters on what exactly she was looking for" (P13).

## 5 DISCUSSION

This study elicited and charted variations in non-experts' thoughts about LLM bad behaviors. From interviews with 28 individuals—with a range of expertise and beliefs about badness—patterns were documented within and across construal levels (more abstract/distal and more concrete/proximal) and types of moral violations. Identification of bad behavior was limited as people responded to abstract elicitations, with relatively few mentions of inaccuracy and nontransparency (which can be linked to popular discourses around AI problematics), and invasiveness in other technologies. With increasingly concrete elicitations (non-valenced request for evaluation of AI, then specific request for thoughts on bad behaviors), characterizations became more numerous and more varied to respectively include poor performance and then two classes of behavior—acts against reality and acts against humans. Some bad behaviors in general comments did *not* carry over to specific evaluations despite being present in the scenario (diminishing human opportunity, invasiveness, resource hogging, animating overreliance). When evaluating very specific AI behaviors through exemplar scenarios, participants did interpret bad behaviors. They identified AI behaviors associated with injury, subterfuge/non-transparency, resistance/aggression/trespass, nonconsensual data use/non-transparency/privacy violation, belligerence/dysfunction/nonsensicalness—all of which are theoretically linked to their inspiring moral-foundations violations of harm, unfairness, subversion, betrayal, degradation, oppression, and dishonesty. Additionally, across all scenarios, participants also characterized poor/absent performance and antisociality (i.e., rudeness or impropriety) as bad behaviors. Also incidentally observed: Participants saw humans as culpable in AI's bad behaviors—commonly the programmers'/developers' roles in making the AI bad but also pointing to users as the guilty parties in poorly prompting, misusing, being silly or stupid, or over-relying on the AI. Altogether, participants characterized AI bad behavior in approximate alignment with moral foundations domains, though antisociality and poor performance were universally panned as problematic (RQ1). As elicitations moved from more general and abstract to more specific and concrete, badness of AI behaviors became more salient and varied (RQ2).

## 5.1 A Tentative Framework for Considering AI Bad Behavior

This study's framing literature review suggested that AI bad behavior can operationally be considered according to two types of badness—functional and moral. The findings, however, suggest a more nuanced framework may be appropriate when considering the *construal* of such behavior by situating interpretations of proximal behaviors within the higher-order types of abstract interpretation (Figure 1). In particular, findings indicate people see AI in terms of two primary moral patients (i.e., victims): There are acts against humans and acts against reality. Some kinds of bad behaviors are more salient for each patient when people consider AI behavior as a distal abstraction compared to the wider possibilities and variations when considering concrete, proximal exemplars. In other words, in line with Construal Level Theory; Trope and Liberman 2010), more salient behaviors were more commonly mentioned early in the interviews, when participants had not yet been prompted to think about bad behavior, likely because they are more accessible as people think about AI (see Young and Fazio 2013).

Among acts against humans, most salient (i.e., accessible at high construal levels) are forms of bad behavior are those associated with oppression and harm—domination over humans, displacing human agency, and causing direct or indirect injury. Among acts against reality, most salient are behaviors enacting dishonesty, degradation, or unfairness—hallucination, deception, bias, producing nonsense, or acts of transparency. Two forms of bad behaviors were primary themes across most of the foundation-based scenarios. Discordance includes the creation of friction or conflict through the violation of norms for appropriate social behavior—the AI was being disingenuous, disrespectful, aggressive, inappropriate, belligerent, severe in ways that degrade human-AI rapport (see Spencer-Oatey 2000). Non-performance includes the AI giving poor quality responses, avoiding or rejecting its job, or doing something other than what was requested; it has in some way deviated from its script and so may be seen as functionally disobedient (see Akrich 1994).

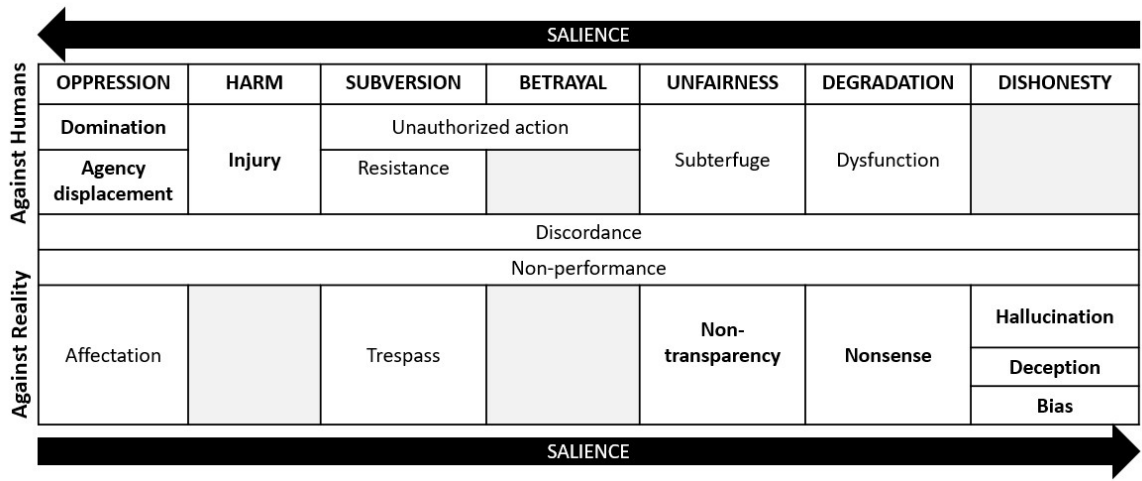

Figure 1: Proposed framework for charting AI bad behavior according to the affected target and moral domain, with discordance and non-performance in common.

This framework comports with the more general notion that AI can be seen as both functionally and morally bad. The more prevalent notions of discordance and non-performance may be seen as violations of the AI's purpose—to be a low-risk, high-benefit support tool for the human's aims and activities (Brauner, et al. 2025) that is generally obedient and moderately congenial (Sun and Wang 2025). The salience of AI behavior badness at different construal levels—across patients and domains—is an important consideration because forms of badness accessible at different levels are likely to have different priming sources and potentially different outcomes. Those salient at high construal levels (domination of and injury to humans and inaccuracy and opaqueness as it represents the world) happen to be aligned with mainstream narratives of human replacement and hallucinations (e.g., Allaham et al. 2025). Those at low construal levels (going into inappropriate territories, behaving without authorization, undermining, and brokenness) may actually be more immediate and concerning to people as they play into individual privacy and functionality domains (e.g., Asthana et al. 2024).

## 5.2 Considering a Sociomorphic Lens

That nonperformance and discordance span nearly the full range of moral-violation scenarios requires critical consideration—they are decidedly *functional* and *social* forms of badness, respectively. Human understanding of nonhuman agency—including AI agencies inherent to the enactment of badness—moves necessarily between the technical and the social (Akrich 1994). We consider the thing, what the thing does, and the outcomes of that doing; the thing is an object (insofar as we know) and the thing it does appears quite social. In the present study attending to the *construal* of AI bad behavior, I only had access to the verbal externalizations as people tried to explain what they saw and believed about it; people do not always have access to the words required to make clear their intuitive experiences and are likely only able to see it through a lens of humanness (see Spatola et al. 2022). The foundation-nonspecific presence of discordance and non-performance suggests those could, together, be a foundational mechanism for the perception of AI badness-in-context.

A productive lens for considering that potential synergy is Seibt and colleagues' (2020) notion of sociomorphism: "the perception of actual non-human social capacities" (p. 51). In contrast to a popular definition of anthropomorphism as an inferenced projection of human qualities onto nonhuman things to rationalize observable behavior (Duffy 2003), Seibt characterizes sociomorphism as a *direct perception* of nonhuman social capacities in nonhuman entities that gives rise to experienced sociality (i.e., feelings of being-with). This can perhaps be best understood by considering a familiar nonhuman—a domestic dog offers social signals through play bows or tail wags. Those are observable behaviors that would not be expected of human yet can create a sense of understanding and being-with. Applied to AI bad behaviors, we can think of it similarly. An AI can convey performance indicators of searching the web, offer immediate responses, generate words on-screen one by one, is sycophantic and overly excited; we would not expect those of humans. If an AI is thought to behave badly when it is discordant and performs poorly, we may think the same of a human, but we would also assume a human assistant would be imperfect. Humans often have bad days, have diverging opinions, slack on the job, make mistakes. In relation to AI, there seems to be a more universal and non-exceptional expectation for congeniality and performance such that they could *together* be signifying some other kind of (yet-unnamed) expected sociomorphic capacity—the absence of which is perceived as bad behavior. For instance, the identified bad behavior of *invasiveness* (i.e., the popping-up of AI functions in unexpected or inappropriate places) might fall into this category—AI is functioning according to its design but doing so in ways that make its presence unexpected and undesired.

## 5.3 Limitations and Future Research

This study is subject to the usual limitations of interview-based, interpretive study designs—a small convenience sample of people answered limited-scope questions about a narrow set of ideas and events that were interpreted by a single analyst. These acknowledged, the posed framework should be considered tentative and must be decomposed and tested to consider whether and how the proposed relations among moral foundations, salience, and focal patients are evident and in varied value systems, populations, and moral operationalizations; theoretically relevant individual differences must also be considered.

Those limitations should be addressed in future research, alongside the other open questions this study has animated and can inform. Perhaps most evidently, there are *gaps* in the proposed framework. Very few participants recognized the betrayal *as betrayal* and instead turned blame to the human; does this suggest they feel AI cannot betray, or might betrayal perceptions unfold in other circumstances? Dishonesty generally manifested as harms against reality—as misrepresentations of the world, as deception, and as manifestations of bias *in principle* and not against humans; does this suggest a non-recognition of the ways inaccuracies could impact people? Has public discourse so framed hallucination as a bad thing-in-itself that its impacts are known but not salient? It is also curious that some mainstream discourses were *not* prevalent in these data—hogging environmental resources, stealing from writers and artists, skills/intellect decay, and lack of accountability were mentioned by only a few participants. How is it that a mainstream narrative that ostensibly *ought* to be part of individual's moral agendas (see Scheufele and Lewenstein 2005) shift into or out of salience? To what extent are these dynamics entangled with the complexity of badness: What counts as an agent and a patient, what counts as badness, was an event intentional, who/what is to blame, what is the outcome, was the outcome deserved, and how was the story of these factors told (e.g., Knobe 2003). With respect to construal levels, outside of manifest considerations like time and space (Trope and Liberman 2010), how do people find AI bad behaviors (with their non-materiality, current low-agency, and malleability) to be self-relevant and proximal? Perhaps most broadly and ambitiously, the proposed framework's core proposition should be explored: The nature of perceived AI morality may lie somewhere in the middle of functional and moral, amid the humanistic and technical, between glitch and perfection.

**Funding**: Research was sponsored by the Army Research Office and was accomplished under Grant Number W911NF-25-1-0079. The views and conclusions contained in this document are those of the authors and should not be interpreted as representing the official policies, either expressed or implied, of the Army Research Office or the U.S. Government. The U.S. Government is authorized to reproduce and distribute reprints for Government purposes notwithstanding any copyright notation herein.

Maninger T, Shank, DB (2022) Perceptions of violations by artificial and human actors across moral foundations. Comp Hum Behav Rep 5:100154. https://doi.org/10.1016/j.chbr.2021.100154

Naaman M (2022) My AI must have been broken: How AI stands to reshape human communication. in Proceedings of RecSys. https://doi.org/10.1145/3523227.3555724

National Institute of Standards and Technology (2023) Artificial Intelligence Risk Management Framework (AI RMF 1.0). U.S Department of Commerce. https://nvlpubs.nist.gov/nistpubs/ai/nist.ai.100-1.pdf

Qureshi NS (2023) Waluigi, Carl Jung, and the Case for Moral AI. Wired. https://www.wired.com/story/waluigi-effect-generative-artificial-intelligence-morality/

Robert LP, Pierce C, Marquis E, Kim S, Alahmad R (2020) Designing fair AI for managing employees in organizations: A review, critique, and design agenda. Hum-Comp Interact 35:545-575. https://doi.org/10.1080/07370024.2020.1735391

Ryazanov I, Öhman C, Björklund J (2024) How ChatGPT Changed the media's narratives on AI: A semi-automated narrative analysis through frame semantics. [preprint] https://arxiv.org/abs/2408.06120

Scheufele DA, Lewenstein BV (2005) The public and nanotechnology: How citizens make sense of emerging technologies. J of Nanopart Res 7:659-667. https://doi.org/10.1007/s11051-005-7526-2

Seibt J, Vestergaard C, Damholdt MF (2020) Sociomorphing, not anthropomorphizing: Towards a typology of experienced sociality. Proc Robophil 2020. IOS Press, pp 51-67.

Shank DB, DeSanti A (2018) Attributions of morality and mind to artificial intelligence after real-world moral violations, Comp Hum Behav 86:401-411. https://doi.org/10.1016/j.chb.2018.05.014

Solman P, Holmes RC (2024) As artificial intelligence rapidly advances, experts debate level of threat to humanity. PBS News Hour. https://www.pbs.org/newshour/show/as-artificial-intelligence-rapidly-advances-experts-debate-level-of-threat-to-humanity

Spatola N, Marchesi S, Wykowska A (2022) Cognitive load affects early processes involved in mentalizing robot behaviour. Sci Rep 12:14924. https://doi.org/10.1038/s41598-022-19213-5

Spencer-Oatey H (2000) Rapport management: A framework for analysis. In: Culturally Speaking: Managing rapport through talk across cultures. Continuum, pp 11-46.

Srdarov S, Leaver T (2024) Generative AI glitches: The artificial everything. M/C J 27(6). https://doi.org/10.5204/mcj.3123

Stenseke J (2024) On the computational complexity of ethics: Moral tractability. Art Int Rev 57:105. https://doi.org/10.1007/s10462-024-10732-3

Sullivan YW, Wamba SF (2022) Moral judgments in the Age of Artificial Intelligence. J of Bus Ethics 178:917-943. https://doi.org/10.1007/s10551-022-05053-w